\documentclass[aps,twocolumn,amsmath,amssymb,showpacs,prl,superscriptaddress]{revtex4}

\usepackage{epsf}

\newcommand{\etal}{{\it et al.}}

\begin{document}

\title{Crossover from coherent to incoherent electronic excitations in the normal 
state of Bi$_2$Sr$_2$CaCu$_2$O$_{8+\delta}$}

\author{A. Kaminski}
\author{S. Rosenkranz}
\affiliation{Department of Physics, University of Illinois at Chicago, Chicago, 
Illinois 60607, USA}
\affiliation{Materials Science Division, Argonne National Laboratory, Argonne, 
Illinois 60439, USA}

\author{H. M. Fretwell}
\affiliation{Department of Physics, University of Wales Swansea, Swansea, SA2 8PP, 
UK}

\author{Z. Li}
\author{H. Raffy}
\affiliation{Laboratoire de Physique des Solides, UniversitŽ Paris-Sud,91405 Orsay, 
France}

\author{M. Randeria}
\affiliation{Tata Institute for Fundamental Research, Homi Bhabha Road, Mumbai 400005, 
India}

\author{M. R. Norman}
\affiliation{Materials Science Division, Argonne National Laboratory, Argonne, 
Illinois 60439, USA}

\author{J. C. Campuzano}
\affiliation{Department of Physics, University of Illinois at Chicago, Chicago, 
Illinois 60607, USA}
\affiliation{Materials Science Division, Argonne National Laboratory, Argonne, 
Illinois 60439, USA}

\date{\today}

\begin{abstract}
Angle resolved photoemission spectroscopy (ARPES) and resistivity 
measurements are used to explore the overdoped region of the high 
temperature superconductor Bi$_2$Sr$_2$CaCu$_2$O$_{8+\delta}$.  We find 
evidence for a new 
crossover line in the phase diagram between a coherent metal phase for 
lower temperatures and higher doping, and an incoherent metal phase for 
higher temperatures and lower doping.  The former is characterized by two 
well-defined spectral peaks in ARPES due to coherent bilayer splitting and 
superlinear behavior in the resistivity, whereas the latter is 
characterized by a single broad spectral feature in ARPES and a linear 
temperature dependence of the resistivity.
\end{abstract}


\maketitle

The normal state of optimal and underdoped high temperature 
superconductors (HTSCs) exhibits anomalous transport and spectroscopic 
properties which have long been recognized as one of the central mysteries 
of the field \cite{1,2}. The electronic excitations are unlike those of 
conventional metals, where one can think of dressed electrons - 
quasiparticles. Instead, the response in the normal state of the HTSCs is 
incoherent, with no identifiable single particle like excitations. The key 
question is how the strange metal evolves into the conventional one at 
high doping. Some models propose that the incoherent normal metal 
represents a new state of matter \cite{3}, with a crossover to more 
conventional behavior \cite{4}. Others suggest that the incoherent state is a 
result of underlying competing interactions, and therefore its behavior 
evolves continuously from the conventional one \cite{5}.

Here, using photoemission and resistivity, we study the issue of the coherence
of the electronic excitations. We find a new crossover line in the phase diagram 
of the HTSCs between the low temperature, overdoped side with coherent 
electronic excitations, and the high temperature, underdoped side, where 
this coherence is lost, and therefore conclude that indeed the two states 
are qualitatively different.

For our measurements, we have chosen Bi$_2$Sr$_2$CaCu$_2$O$_{8+\delta}$ 
(Bi2212), which has 
the advantage that it can be prepared over a wide range of doping, and is 
ideal for angle resolved photoemission (ARPES) given its quasi two 
dimensionality. Additionally, it possesses two $CuO_2$ layers (a bilayer), 
and thus the issue of coherence can be probed by not only examining its 
planar properties as a function of doping and temperature, but also by 
looking for the presence of bilayer splitting; if the motion of electrons 
within the bilayer is coherent, then we expect the formation of 
antibonding (A) and bonding (B) states.  These are the antisymmetric and 
symmetric combinations of the layer wavefunctions with energies 
$\epsilon_{A(B)}({\bf k})=\epsilon({\bf k}) \pm t_{\perp}({\bf k})$ where 
$\epsilon({\bf k})$ is the planar dispersion and $t_{\perp}({\bf k})$
the interplanar coupling. As the $CuO_2$ planes in 
Bi2212 are separated by only 3.17\AA, electronic structure calculations 
predict a sizable bilayer splitting of order 0.3 eV \cite{6}. In the coherent 
regime, the primary effect of interactions would be to renormalize the 
splitting to a smaller value, without qualitatively affecting the 
spectrum.  On the other hand, if the interactions are sufficiently strong, 
then we expect the coherent behaviour within each plane, as well as the 
coherent motion within the bilayer, to be destroyed \cite{1}. The in-plane 
effect would be reflected in both the ARPES spectral lineshape and in the 
temperature dependence of the planar transport. The out-of-plane effect 
would be a loss of the coherent bilayer splitting.

The ARPES and resistivity data were obtained on high quality epitaxially 
grown thin films of Bi2212 \cite{7}. These films have the very useful feature 
of displaying small signals from the structural superlattice distortion
($<$ 3\%), which otherwise would complicate \cite{8} the interpretation of ARPES 
data in the vicinity of the  $(\pi,0)$ point of the Brillouin zone. This is 
particularly important for us, since bilayer splitting is maximal at $(\pi,0)$ 
\cite{9,10,11,12}, as shown in panel a of Fig.~1.

\begin{figure}
\centerline{\epsfxsize=2.4in{\epsfbox{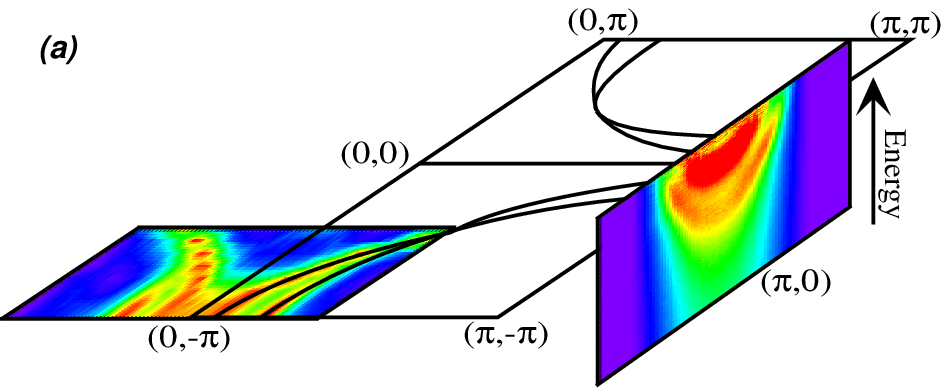}}}
\centerline{\epsfxsize=3.2in{\epsfbox{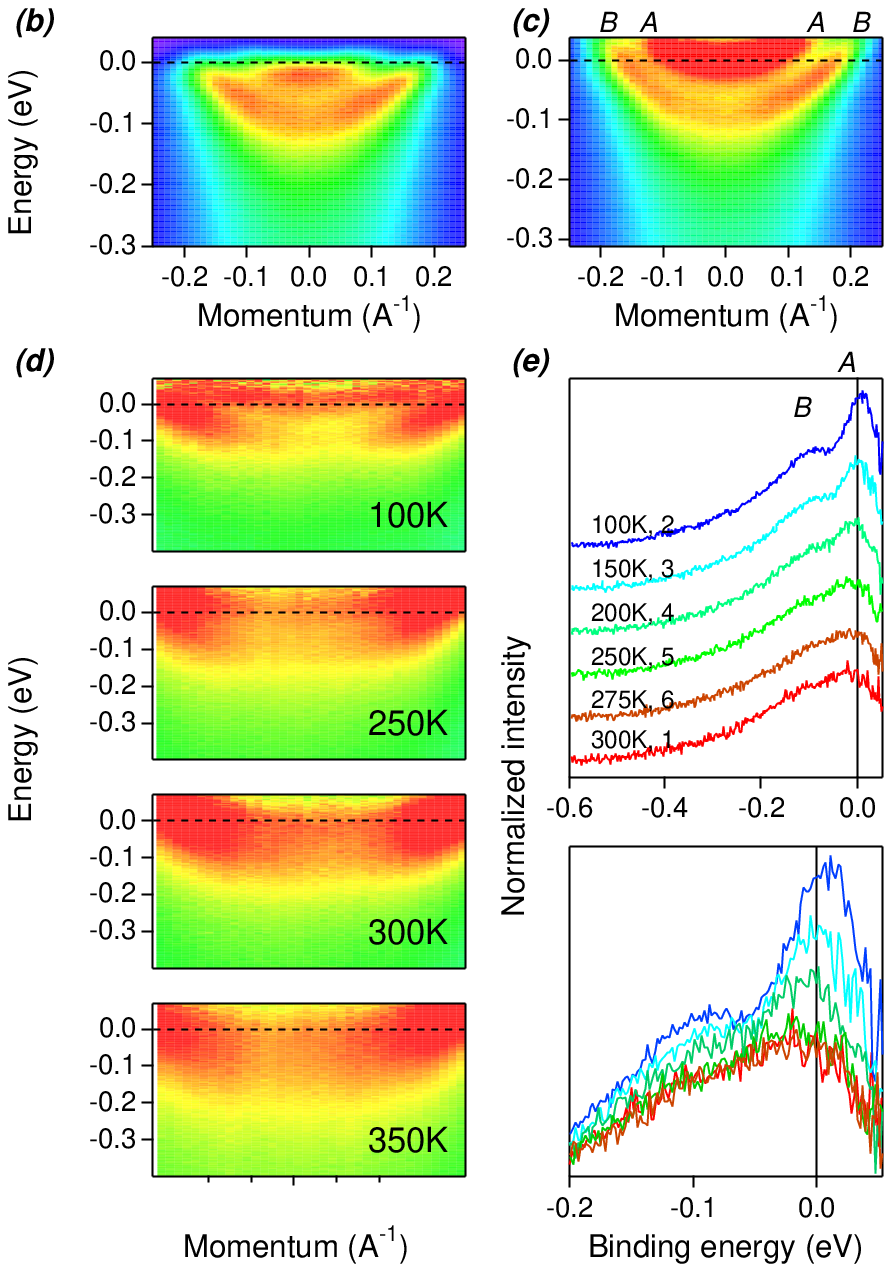}}}
\caption{
ARPES data for overdoped ($T_c$ = 52K) Bi2212 samples. (a) Intensity 
versus momentum and energy for T=100K. The two solid curves represent the 
bilayer split Fermi surface. (b) Intensity for momenta along 
$(\pi,0)-(\pi,\pi)$ at T=100K, 
with plots centred at $(\pi,0)$. (c) The same data divided by the Fermi 
function. (d) Same as (c), but at various temperatures.
(e) Spectrum at $(\pi,0)$ (divided by the 
Fermi function) at various temperatures. All curves are overlapped on the 
bottom of the panel to demonstrate lack of temperature dependence of the 
lineshape above 250K.
}
\label{fig1}
\end{figure}
ARPES measurements were carried out
at the Synchrotron Radiation Center in 
Wisconsin with an energy resolution of 30 meV and a momentum resolution of 
0.01\AA using our SES50 analyzer and undulator 4 meter NIM beamline. The 
ARPES intensity as a function of the planar momentum {\bf k} and energy $\omega$
(measured with respect to the chemical potential) is given by \cite{13}
$I({\bf k},\omega)=I_0({\bf k})f(\omega)A({\bf k},\omega)$
(convolved with the resolution function). Here, $I_0$ is an intensity 
prefactor, $f$ the Fermi function, and $A$ the single particle spectral 
function, which measures the probability of removing or adding an electron 
from the system. The peak in $A({\bf k},\omega)$ measures the energy of the 
electronic 
excitation, while its linewidth is inversely proportional to the lifetime. 
Since the ARPES lineshape is ``chopped off" by the Fermi function, we 
choose in some cases to divide our data by a resolution broadened Fermi 
function, obtained by fitting the leading edge of a polycrystalline Au  in 
contact with the sample. This procedure allows us to focus directly on
$A({\bf k},\omega)$
(albeit approximately because of the resolution convolution).

\begin{figure}
\centerline{\epsfxsize=3.2in{\epsfbox{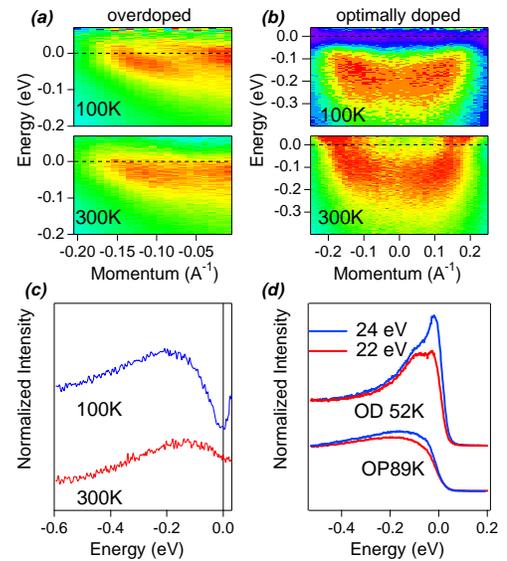}}}
\caption{
ARPES data for samples at various dopings. (a) As in Fig.~1(d), 
but for an overdoped ($T_C$ = 75K) sample. (b) As in Fig.~1(d), but for an 
optimal doped ($T_C$ = 89K) sample  (at 100K the intensity at the chemical 
potential is suppressed due to the pseudogap). (c) Spectrum at $(\pi,0)$
(divided by 
the Fermi function) for an optimal doped ($T_C$ = 89K) sample. (d) Raw data 
at $(\pi,0)$ at two different photon energies for an overdoped ($T_C$ = 52K 
sample) 
and an optimal doped ($T_C$ = 89K) sample at T= 100K.
}
\label{fig2}
\end{figure}
In panel (b) of Fig.~1 we plot raw ARPES data for an overdoped (OD) sample 
($T_C$ = 52K) at T=100K, along a momentum cut centred at the $(\pi,0)$
point of the 
Brillouin zone. In addition, in panel (c), we plot the same data divided by 
the Fermi function, which approximates the true spectral function. The 
data in panel (c) reveal two dispersing bands due to the bilayer splitting, 
with the A band close to, and the B band well below, the chemical 
potential. In panel (d), we show data like that in (c) (divided by the Fermi 
function) for another sample with the same $T_C$ as a function of 
temperature. The bilayer splitting can clearly be seen at 100K, however 
above 250K the two bands are no longer observed. To obtain more precise 
information, in panel (e) we show the temperature dependence of the spectral 
function at $(\pi,0)$ as a function of energy (raw data divided by the Fermi 
function). The sample was temperature-cycled when taking the data to 
ensure that the observed effect is intrinsic and not due to the sample 
aging (the numbers in the legend indicate the order of measurement). At 
100K, one sees clearly the presence of two peaks, a sharp A peak near the 
chemical potential, and a broader B peak at about 100 meV below. As the 
temperature is increased, the peaks broaden and lose intensity, until only 
a single broad peak remains at 250K. At the bottom of the panel (e), we plot 
the curves for all temperatures without an offset to show that lineshape 
changes occur only up to 250K. Based on this, we argue that above 250K the 
system no longer exhibits coherent excitations, both in regards to inverse 
lifetime (spectral peak widths) and bilayer splitting (appearance of two 
separate spectral peaks).  That is, the data indicate that {\it both} 
in-plane and out-of-plane coherence are lost.

\begin{figure}
\centerline{\epsfxsize=3.4in{\epsfbox{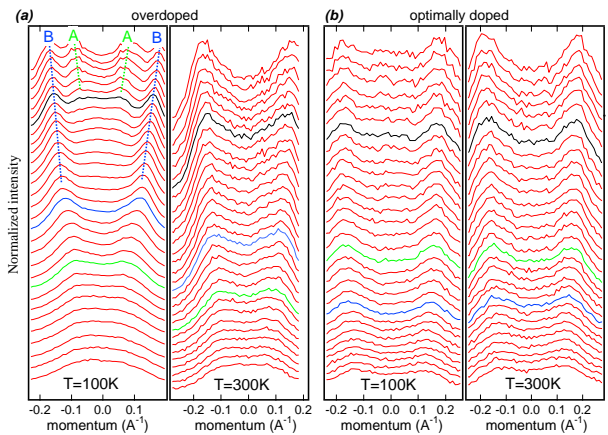}}}
\caption{
Momentum distribution curves (MDCs) along $(\pi,0)-(\pi,\pi)$ for various 
energies. 
The black curves are at the chemical potential, the blue at -50 meV and 
the green at -100 meV. The curves for the energies above -50 meV are 
spaced every 5 meV and remaining curves every 10 meV. (a) Overdoped ($T_C$ = 
52K) sample at 100K (``A'' and green dotted lines mark antibonding peaks, 
``B'' and blue dotted lines - bonding peaks) and 300K. (b) Optimal ($T_C$ = 
89K) sample at 100K and 300K. The data are quite noisy at 100K close to 
the chemical potential because of the intensity being suppressed due to 
the pseudogap.
}
\label{fig3}
\end{figure}
We now examine the issue of coherence as a function of doping. In panel (a) 
of Fig.~2, we show data like in Fig.~1(d), but for an overdoped ($T_C$ = 75K) 
sample. Again, note the presence of bilayer splitting at 100K which is not 
visible at 300K. We can contrast this behaviour with that of an optimally 
doped sample ($T_C$ = 89K) shown in panel (b), where the intensity plots do not 
indicate the presence of bilayer splitting, even at 100K. This is further 
illustrated in panel (c), where again the spectrum at $(\pi,0)$ (divided by the 
Fermi 
function) is shown. At 100K, only a single broad peak is seen, with no 
presence of bilayer splitting, indicating incoherent behaviour. Instead, a 
pseudogap is seen, centred at the chemical potential, which fills in as 
the temperature is increased. An important check can be made by analyzing 
the photon energy dependence of the data. It has been recently observed 
that the spectral lineshape changes as a function of photon energy for 
overdoped samples due to the relative weighting of the A and B peaks 
\cite{11,12}. This is clearly seen in Fig.~2(d), where data at $(\pi,0)$
for the overdoped 
sample of Fig.~1 is shown for two different photon energies. In contrast, 
for the optimal doped sample, only a very small change with photon energy 
is observed (panel d), indicating the absence of bilayer splitting.

It is also useful to plot the data as a function of momentum for fixed 
energy (momentum distribution curve, or MDC). In Fig.~3 we show such plots 
for a few values of the binding energy, at low and high temperatures, for 
two samples, an overdoped sample like that in Fig.~1, and another optimal 
doped sample ($T_C$ = 89K). At 100K and energies close to the chemical 
potential, the overdoped MDC has four peaks (panel a). The two peaks 
closest to the centre of the plot correspond to the A band, while the two 
peaks on the outside correspond to the B band. As the binding energy 
increases, the two A peaks approach each other and then merge into a 
single peak, which then disappears at still higher energies. This is the 
expected behaviour of MDCs close to the bottom of a band, and is also 
observed for the B peak at even higher binding energies. However, at high 
temperatures only two peaks are observed, regardless of binding energy 
(panel a). This can be contrasted with the optimally doped sample, where 
only two peaks are visible in the MDCs, whether at 100K or 300K (panel b).

\begin{figure}
\centerline{\epsfxsize=3.2in{\epsfbox{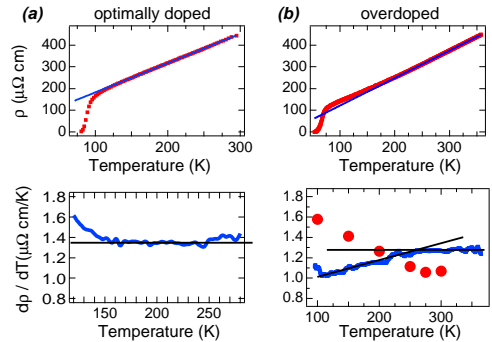}}}
\caption{
Resistivity data. The blue lines denote linear T fits to the 
high temperature data for (a) optimal doped ($T_C$ = 89K) and (b) overdoped ($T_C$ 
= 52K) samples (top graphs). The bottom plots are the temperature 
derivative of the resistivity for the two dopings, with black lines guides 
to the eye.  Red dots on bottom right plot are the ARPES intensities from 
Fig.~1(e).
}
\label{fig4}
\end{figure}
We now connect our ARPES observations with transport data taken on the 
same films. The resistivity was determined using the standard four-probe 
method. In Fig.~4 we plot the resistivity as a function of temperature for 
the optimally doped (panel a) and overdoped (panel b) samples of Fig.~3, 
with the black line a linear fit to the high temperature data. Both 
samples at high temperatures exhibit a linear T resistivity, which has 
been linked to the absence of coherent quasiparticles \cite{1,2}. For the 
optimally doped sample, this behaviour continues to near $T_C$, with the 
rounding just above $T_C$ due to fluctuation effects. In contrast, the 
overdoped sample shows strong deviations from linearity, which set in at 
about 250K. These results can be understood more easily by plotting the 
derivative of the resistivity, which emphasizes the strength of the 
inelastic scattering contribution. One can see that while the optimally 
doped sample shows a constant derivative, the overdoped sample shows a 
change in derivative at 250 K. Below this temperature, the derivative 
monotonically decreases. The superlinear behaviour of the resistivity 
below 250K indicates the presence of coherent excitations. We note the 
strong correlation of these observations with ARPES. In the optimally 
doped sample, sharp spectral peaks only begin to appear at temperatures 
slightly above $T_C$. And in the overdoped sample (Fig.~1e), the sharp A peak 
disappears above 250K. As we show in the bottom right panel of Fig.~4, 
the ARPES intensity is inversely related to $d\rho/dT$. In particular, the 
intensity becomes constant when the resitivity becomes linear.

Based on our results, we show in Fig.~5 a proposed phase diagram for the 
HTSCs. The crossover between the pseudogap phase and the strange metal 
phase has been studied in the past, both by ARPES \cite{14} and 
transport \cite{7}. 
What we have shown here is the presence of a new crossover line, between a 
conventional metal phase on the overdoped side of the phase diagram and 
the strange metal phase, seen from both spectroscopic and transport 
measurements. The two crossover lines (pseudogap/strange metal and 
metal/strange metal) were determined by the departure from linear T 
resistivity. In all cases, the crossover lines were verified by ARPES, 
with the pseudogap line being determined by the closing of the leading 
edge gap at $(\pi,0)$, and the metal line by the loss of the sharp (A) peak
at $(\pi,0)$.

\begin{figure}
\centerline{\epsfxsize=2.4in{\epsfbox{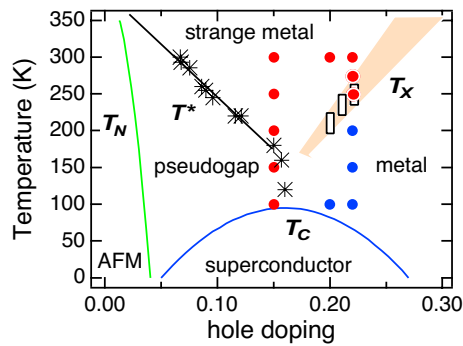}}}
\caption{
Phase diagram for the HTSCs. The strange metal/pseudogap 
transition (black stars) and strange metal/metal crossover (black squares) 
are obtained from the departure from linear T resistivity. The open red 
circles correspond to ARPES data where bilayer splitting was not observed, 
the blue dots correspond to ARPES data showing bilayer splitting.
}
\label{fig5}
\end{figure}
The crossover line we have found from the loss of coherence has been long 
predicted on theoretical grounds. Slave boson studies of the t-J model 
\cite{4,15} predict a phase diagram very similar to Fig.~5, with the crossover 
line between the strange metal and metal phases marking the ``condensation 
of holons" (i.e., for temperatures below this, the doped holes have phase 
coherence).  However, a similar crossover to the one observed here may 
also be expected near a quantum critical point \cite{16}, with the ``ordered" 
region corresponding to the pseudogap, the disordered region the 
conventional metal, and the quantum critical regime the strange metal 
phase. Further studies are needed to distinguish these possibilities.

In conclusion, our data show the presence of a coherent normal 
metal in overdoped samples, and that this state crosses over into an 
incoherent metal at higher temperatures.  We emphasize that the 
ARPES data indicate a loss of {\it both} in-plane and out-of-plane
coherence.  Furthermore, this crossover temperature increases with 
doping.
Our studies indicate that the normal state of the HTSCs, with its various 
phases, is a much richer field of study than even its exotic 
superconducting state, and has strong implications for the many body 
theory of electrons in reduced spatial dimensions.

This work was supported by the NSF 
DMR 9974401 and the U.S. DOE, Office of Science, 
under Contract No. W-31-109-ENG-38. The Synchrotron Radiation Center is 
supported by NSF DMR 9212658.
SR was supported in part by the Swiss National Science Foundation, 
and MR by the Indian DST through the Swarnajayanti 
scheme.  We acknowledge helpful discussions with P.D. Johnson.

\end{document}